\documentclass[preprint,12pt]{elsarticle}

\usepackage{graphics}
\usepackage{graphicx}

\usepackage{amsmath}
\usepackage{amsthm}
\usepackage{epstopdf}
\usepackage{dcolumn}
\usepackage{bm}
\usepackage{color}
\usepackage[
total={6.5in,8.75in}, top=1.5in, left=0.9in, 
]{geometry}

\biboptions{comma,sort,compress}

\journal{Physica A}

\begin{document}

\begin{frontmatter}



\title{Theory of earthquakes interevent times applied to financial markets}

\author[1,2]{Maciej Jagielski\corref{cor1}}
\ead{zagielski@interia.pl}
\cortext[cor1]{Corresponding author. Tel.: +41 44 632 82 58}
\author[1]{Ryszard Kutner}%
\author[2]{Didier Sornette}

\address[1]{Faculty of Physics, University of Warsaw, Pasteura 5, PL-02093 Warszawa, Poland}
\address[2]{ETH Z\"urich, Department of Management, Technology and Economics, Scheuchzerstrasse 7, CH-8092 Z\"urich, Switzerland}

\begin{abstract}
We analyze the probability density function (PDF) of waiting times between financial loss exceedances.
The empirical PDFs are fitted with the self-excited Hawkes conditional Poisson process with a 
long power law memory kernel. The Hawkes process is the simplest extension of the Poisson process that takes into account how past events influence the occurrence of future events. By analyzing the empirical data for 15 different financial assets, we show that the formalism of the Hawkes process used for earthquakes can successfully model the PDF of interevent times between successive market losses.  
\end{abstract}


\end{frontmatter}


\section{Introduction}\label{intro}

The activity of many complex systems in social and natural sciences can be characterized by sporadic bursts followed by long periods of low activity. To quantitatively describe this kind of dynamics, we can use the interevent times (also called ``pausing time", ``waiting time", ``intertransaction time", ``interoccurrence time", and ``recurrence time"), defined as times between two consecutive events of high (suitably defined) activity of the system. It was shown in many research papers that the probability density distribution (PDF) of interevent times plays the role of a universal characteristic of dynamical complex systems. Examples include anomalous transport \cite{ZT_1991}, pattern discovery \cite{GCH_2005}, earthquakes \cite{C_2003,C_2004a,LTHB_2005,SS_2006,SS_2007,C_2004b, BCDS_2002} and rock fractures \cite{DSD_2007}, extreme events and long-memory processes \cite{SK_2008}, email communications \cite{B_2005,MSMA_2008}, web browsing, library visits, stock trading \cite{VODGKB_2006}, human dynamics \cite{B_2005,VODGKB_2006,OV_2009,MSFDS_2011}, social dynamics \cite{RFFMV_2010,CFL_2009}, finance risks \cite{RZ_2010,GKF_2014,CDM_2005,CM_2012}, letter correspondences \cite{MSCA_2009}, and financial markets \cite{LTB_2011,LB_2014,BB_2008,BB_2009,DJGKS_2016}. 

In this paper, we analyze the probability density function of interevent intervals between times when market returns are producing excessive losses. Empirical market data on excessive losses are defined as losses below some negative threshold $-Q$ (or above threshold $Q$ for the absolute value of negative returns). In addition, the mean interevent time can be used as a control variable. The empirical distribution of interevent times when scaled by the mean interevent time is found to be empirically close to a universal statistical quantity unaffected by time scale, type of market, asset, or index.

Denys et al. \cite{DJGKS_2016} used the continuous-time random walk formalism to model
the dependence of the successive interevent times of losses and gains exceedances.
They analytically derived a class of ``superstatistics" that accurately model empirical market activity data 
at different transition thresholds. They measured the interevent times between excessive losses and excessive profits, and used the mean interevent time as a control variable to derive a universal description of empirical data
in the form of an explicit closed form of the probability density function of interevent times.  This function is a power law corrected by the lower incomplete gamma function: it is thus close to an exponential for small values
of the variable and then crosses over asymptotically to a power law tail. 
The class of superstatistics mentioned above for the interevent times is a single-variable statistics, which 
can be viewed as a projection of the bivariate Weibull copula that describes the dependence 
between subsequent interevent times (see Sec. III C in ref. \cite{DJGKS_2016} for details). 
The existence of an interdependence between interevent times, previously 
investigated only on pairwise successive interevents, motivates us to ask whether 
more distant interdependence may be at work. In other words, we want to investigate
the value of the non-Markovian property of the self-excited Hawkes conditional Poisson process, the simplest extension of the Poisson process that takes into account how all past events influence the occurrence of all future events, 
to account parsimoniously for the distribution of interevent waiting times.
Specifically, we reinterpret the model proposed by Saichev and Sornette and collaborators \cite{SS_2006,SS_2007,SUS_2008,SS_2013} to describe the empirical distribution of recurrence times between earthquakes and verify its application to the empirical probability density function of interevent times between market returns exceeding fixed thresholds.


The paper is organized as follows. Section 2 contains a short description of the theoretical approach based on \cite{SS_2006,SS_2007,SUS_2008,SS_2013}. In Section 3, we provide the data description and discuss the obtained results. Section 4 includes our concluding remarks and a summary of the paper.

\section{Theoretical approach}

The self-exciting Hawkes process is described by the following conditional intensity function \cite{H_1971,SUS_2008,SS_2013,FS_2012,FBMS_2014}
\begin{equation}
\lambda(t|H(t))=\mu(t)+\sum_{t_i<t}h(t-t_i).
\label{rown6}
\end{equation}
The conditional intensity is defined such that $\lambda(t)dt$ is the expected number of events in the time interval $[t,t+dt]$. It depends not only on time $t$ but also is conditioned on the history of past events $H(t)=\{\ldots (t_i,m_i),\ldots,t_1,m_1)\}$. Here $t_i<t$ ($m_i<m$) is the time (``mark'') of an event $i$ counted from the present time $t$. The term $\mu(t)$ is the background intensity that accounts for exogenous events (not dependent on history), and $h(t)$ is a memory kernel function that weights how much past events influence the generation of future events and thus controls the amplitude of the endogenous feedback mechanism. The conditional intensity uniquely determines the distribution of the process \cite{GKF_2014}. 


Here we assume that the background intensity is constant, i.e. $\mu(t)=\omega$. In addition, the Hawkes model takes into
that any event can trigger its own progeny over multiple generations \cite{SHSorpdf_2005}.
The average number of first generation events (i.e. directly generated) at time t of a given event occurring at $t_i$ is 
given by  $h(t-t_i)=n\Phi(t-t_i)$, where $n$ is a branching ratio \cite{SS_2007,SUS_2008,FS_2012,FBMS_2014}
\begin{equation}
n\equiv\int_{0}^{\infty}h(t)dt ~,
\label{rown11}
\end{equation}
and $\Phi(t)$ is the normalized memory kernel. Hence, Equation (\ref{rown6}) takes the form
\begin{equation}
\lambda(t|H(t))=\omega+n\sum_{t_i<t}\Phi(t-t_i).
\label{rown7}
\end{equation}

The branching ratio $n$ is the key parameter of the Hawkes model and can also be
interpreted as the fraction of triggered events (or any generation) to the total number of events 
\cite{HS_2003,FS_2012}. The values of the branching ratio define three regimes: (i) $n<1$ -- subcritical, (ii) $n=1$ -- critical, and (iii) $n>1$ -- supercritical. In the regimes (i) or (ii), starting from a single event at time, the process dies out with probability $1$. In regime (iii), there is a finite probability that the process will explode 
on average exponentially to an infinite number of events. 
In the subcritical regime $(n\leq 1)$, the Hawkes process is stationary in the presence of a
constant background intensity $\mu(t)=\omega =$ constant.

The normalized memory kernel $\Phi(t)$ is taken under the form
of the Omori-Utsu law describing the rate (decaying as a power law) of triggered events of first generation from a given event. Denoting the usual Omori-Utsu power law exponent by $p=1+\theta$ and assuming $\theta>0$ ($p>1$), the normalized probability density function of the Omori-Utsu  can be written as \cite{SS_2006,SS_2007,SUS_2008,SS_2013}
\begin{equation}
\Phi(t)=\frac{\theta t^{\theta}_0}{\left(t_0+t\right)^{1+\theta}}\quad 0<\theta   \quad t_0>0, \quad t>0.
\label{rown3}
\end{equation}
The parameter $t_0$ is a characteristic microscopic time scale of the Omori-Utsu  law that ensures regularization at small time and normalization \cite{SS_2013}. 
$\Phi(t)$ is nothing but the PDF of the interevent times between the parent event and the 
offspring events directly triggered by it  (first generation events). 
The power law form (\ref{rown3}) derives from the ``theory of procrastination'' based
on priority queuing theory \cite{GLpriorQue_2006,GLpriorQue_2008,SSorProcras_2009}, 
which predicts $\theta =0.5$. In social systems, $\theta$ is empirically found to be $\theta=0.3-0.4$
\cite{JohSorURL00,DeschaSor04,DeschaSor05,SorRevendo05,CS_2008}. 
The Omori-Utsu law is also valid for financial markets \cite{PWHS_2010,S_2012,HardimanBou13}.

In a series of papers \cite{SS_2006,SS_2007,SUS_2008}, Saichev and Sornette derived
the theoretical expression of the PDF of interevent times, as predicted by the Hawkes process.
The PDF $P(\tau)$ should take the form
\begin{equation}
P(\tau)\simeq \lambda f(\lambda \tau),
\label{rown9}
\end{equation}
where $\lambda$ is the average rate of events and 
 \begin{equation}
f(\lambda \tau)=\left[\frac{n\theta t_0^{\theta}}{\lambda \tau^{1+\theta}}+\left(1-n+\frac{nt_0^{\theta}}{\tau^{\theta}}\right)^2\right]\exp\left[-\left(1-n \right) \lambda\tau-\frac{n\lambda t_0^{\theta}}{1-\theta} \tau^{1-\theta}\right]~.
\label{rown10}
\end{equation}
Recall that $\theta$ and $t_0$ are defined in Eq. (\ref{rown3}), and $n$ is defined by Eq. (\ref{rown11}). 
Our goal is now to test how well expression (\ref{rown9}) with (\ref{rown10})
account for the PDF of interevent times between market returns exceeding fixed thresholds.

\section{Empirical results}

We constructed the empirical probability distribution function of interevent times (for daily data) between excessive losses for 15 representative financial records -- IBM (1962-2010), Boeing (BA; 1962-2010), General Electric (GE; 1962-2010), Coca-Cola (KO; 1962-2010),  Dow-Jones (DJI; 1985-2010),  Financial Times Stock Exchange 100 (FTSE; 1984-2010), 
NASDAQ (2002-2010), S\&P~500 (1950-2010), Brent Crude Oil (BCO, 1987-2010), West Texas Intermediate (WTI, 1986-2010), USD/DKK exchange rate (1971-2010), GBP/USD exchange rate (1971-2010), USD/JPY exchange rate (1971-2010), USD/CHF exchange rate (1971-2010), and EUR/USD exchange rate (1999-2010).

\begin{figure}[!ht]
\centering 
\includegraphics[scale=0.50]{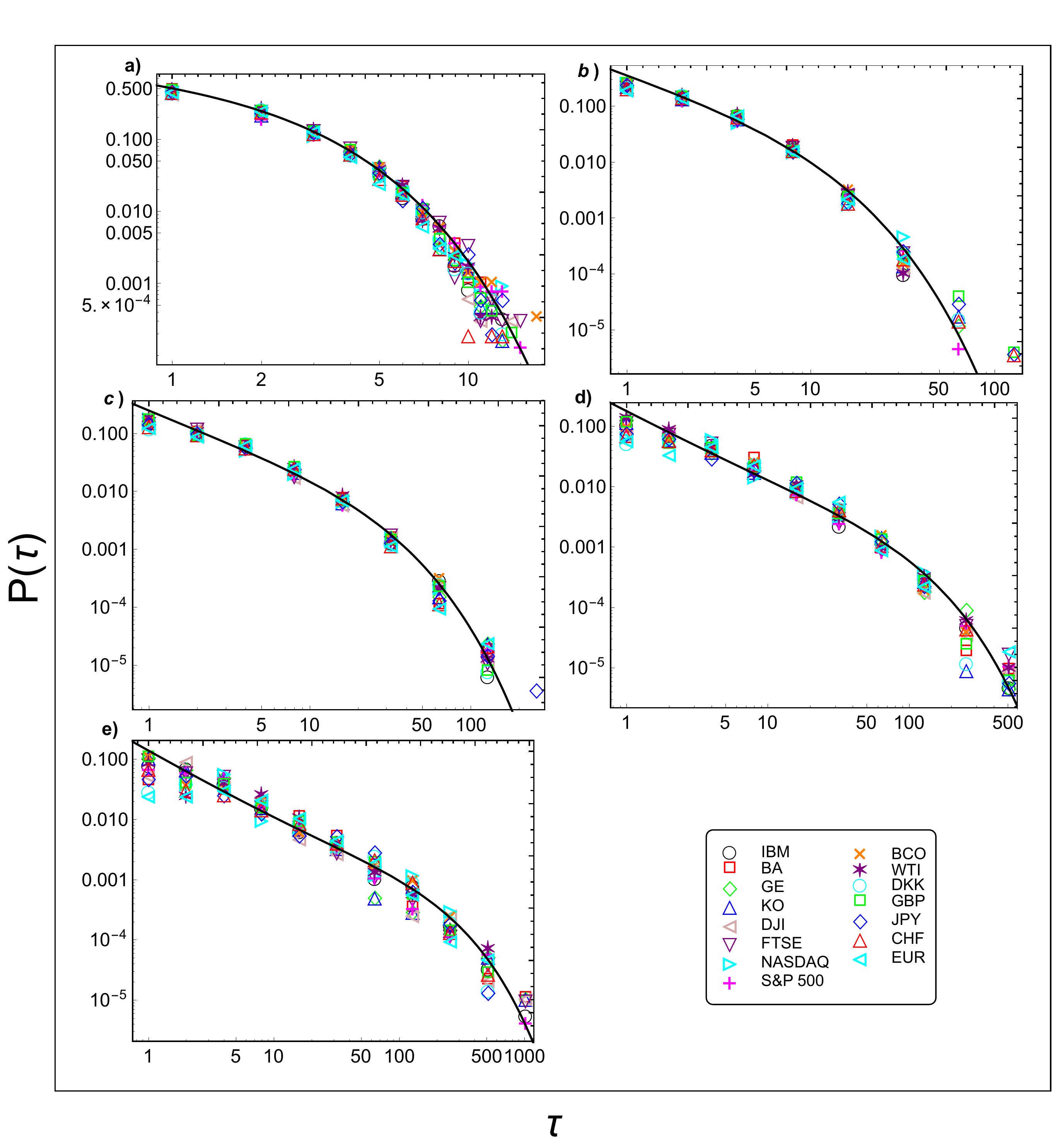}
\caption{Fit of the formula (\ref{rown10}) to the empirical data -- IBM, BA, GE, KO, DJI, FTSE, NASDAQ, S\&P~500, BCO, WTI, USD/DKK, GBP/USD, USD/JPY, USD/CHF, and EUR/USD. Fits and the empirical data are presented for the following mean interevent times (in units of days) a) $\left< \tau\right>=2$, b) $\left< \tau\right>=5$, c) $\left< \tau\right>=10$, d) $\left< \tau\right>=30$, e) $\left< \tau\right>=70$. The plots are in double logarithmic scale.}
\label{fig1}
\end{figure}

Fig. \ref{fig1} shows the fit of formula (\ref{rown10}) to empirical data for $5$ different thresholds. We set the thresholds in such way that the mean (average) interevent time $\left< \tau\right>$ for the empirical distributions is equal to $2$, $5$, $10$, $30$, and $70$ days, respectively. For every financial asset, we calibrated the parameters using the maximum likelihood method. Then, for a fixed mean interevent time, we estimated each parameter by taking an average of its values over the 15 financial records mentioned above. Fig. \ref{fig1} shows the obtained fits with formula (\ref{rown10}) for those averaged parameters\footnote{For a more stable parameter estimation, we considered in formula (\ref{rown10}) the parameter $t_0^{\theta}$ instead of $t_0$.}. The values of the calibrated parameters can be found in Table \ref{tab1}. 

\begin{table}[!ht]
\centering
\caption{Values of the parameters in the formula (\ref{rown10}) calibrated to the empirical data using
the maximum likelihood method. $\left< \tau\right>$ is the mean intervent time. The errors bars of the parameters fall in the range: $n:0.01-0.08, \theta: 0.005-0.1, \lambda: 0.005-0.15, t_0^{\theta}: 0.01-0.03$.}
\begin{tabular}{|c||c|c|c|c|} \hline

$\left< \tau\right>$ & $n$ & $\theta$ & $\lambda$ & $t_0^{\theta}$\\ \hline \hline

2 & 0.78 & 0.077 &	0.899&	0.61\\ \hline

5 & 1.00	& 0.290 &	0.737 &	0.41\\ \hline

10 & 1.00 & 0.267 &	0.462 &	0.29 \\ \hline

30 & 1.00 & 0.250 &	0.254 &	0.15 
\\ \hline

70 & 1.00 &	0.203 &	0.154 &	0.11 
\\ \hline

\end{tabular}
\label{tab1}
\end{table}

Formula (\ref{rown10}) is thus found to describe very well the empirical distributions of interevent times for different mean interevent times. For mean interevent times $\left< \tau\right>$ larger than $2$ days,
we find that the branching ratio $n$ saturates to its upper bound $1$ (imposed
to ensure stationarity). Such a large value of the branching ratio implies that all generations
from first-generation, second-generation up to an infinite number of generations are contributing
to the observed sequences of loss exceedances \cite{SHSorpdf_2005,SSorGen_2010}.
The empirical data is compatible with the diagnostic that, measured via the 
distribution of interevent loss exceedances, financial systems may be in or close to a critical regime \cite{HardimanBou13}
(see however \cite{FS_2015}).
 In any case, our findings confirm the very strong non-Markovian nature of the series of loss exceedances.

Examining the results in more details, one can see that parameters $n$ and $\theta$ are approximately universal for all mean interevent times (except $\left< \tau\right>=2$) while $\lambda$ and $t_0^{\theta}$ are time-scale specific. 
For all thresholds except the lowest one, $\theta$ is in the range $0.2$--$0.3$, 
which is reminiscent of the value $\simeq 0.3$ found for other social systems   \cite{DeschaSor04, CS_2008}. 

\begin{figure}[!ht]
\centering 
\includegraphics[scale=0.50]{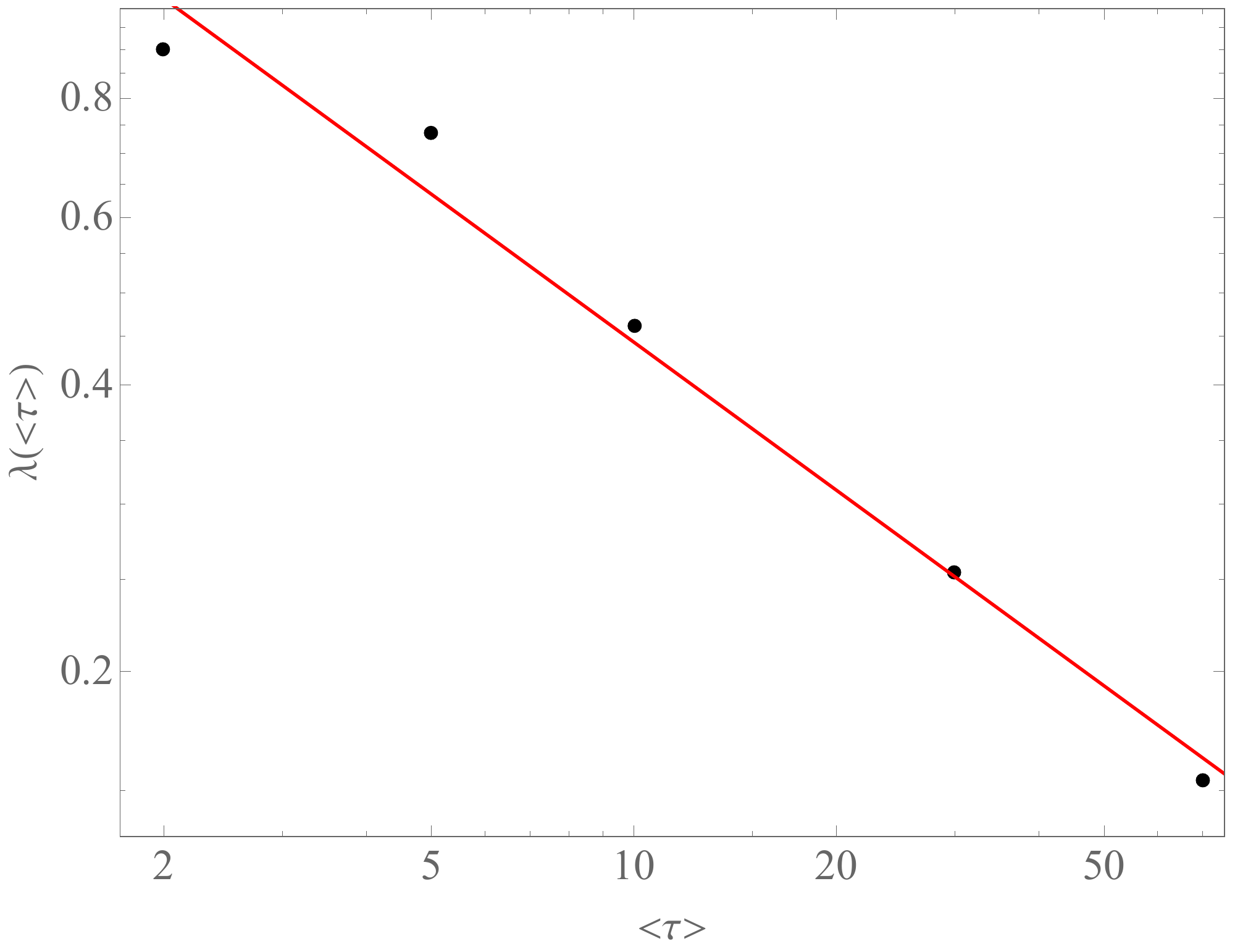}
\caption{Fit of the formula (\ref{rown12}) to the empirical data with the following parameters: $a=0.17\pm 0.04$, $\tau_0=10\pm 2$ and $b=0.90\pm 0.07$. The plot is presented in a log-log scale and the mean interevent times are in units of days. In the presented range of variables formula (\ref{rown12}) imitates a spurious power law.}
\label{fig2}
\end{figure}

If the exceedance losses occurred on time stamps exactly described by the Hawkes process,
the rate $\lambda$ of events should be just the inverse of the average interevent waiting time $\left< \tau \right>$.
In Table \ref{tab1}, one can see that this prediction is violated.  Indeed, changing
the threshold of loss exceedance such that $\left< \tau \right>$ is increased from $2$ to $5$
decreases $\lambda$ by 18\% (and not by a factor of $2.5$). The dependence of $\lambda$ 
as a function of $\left< \tau \right>$ is quite accurately captured by the following empirical formula
\begin{equation}
\lambda\left(\left< \tau\right>\right)=a+b\exp\left(-\frac{\left< \tau\right>}{\tau_0}\right),
\label{rown12}
\end{equation}
with $a=0.17\pm 0.04$, $\tau_0=10\pm 2$ and $b=0.90\pm 0.07$, as shown in 
Fig. \ref{fig2}. This initially weak dependence of $\lambda$ 
as a function of $\left< \tau \right>$ for the smaller values of $\left< \tau \right>$
followed by an accelerated decrease for larger $\left< \tau \right>$'s, is the signature
of a stronger volatility clustering than even described by the self-excited Hawkes process.
Volatility clustering and intermittency are well known properties of financial
returns \cite{ArneoMuzySor98}, which have been previously shown to be best described by multifractal processes
\cite{BacryMuzy01,MuzySorArn01,CalvetFisher1,SorMalMuzy03,CalvetFisher2}.
While improving on previous approaches based on a Markovian or quasi-Markovian
approximation \cite{DJGKS_2016} by accounting for the influence of all past events,
the Hawkes process is only mono-fractal, which thus fails to fully account
for the multi-scale multifractal nature of financial time series.
While a growing literature has been using the Hawkes process to model financial time series,
our present investigation shows both its good properties (long-range self-excitation)
and its limits (mono-fractality). Hence, the controversy on whether financial markets
are critical \cite{HardimanBou13} or not \cite{FS_2012,FS_2015}, 
when examined through the lenses of the Hawkes process, may be 
an artefact of its mono-fractal nature. The real structure of financial returns is likely 
to be much richer and complex, not to speak of the importance of accounting
for regime shifts and even non-stationarity as a results of the influence of major
economic shocks. Perhaps, multifractal extensions of the Hawkes process 
in the spirit of Filimonov and Sornette's ``self-excited multifractal dynamics'' \cite{FilimSor11}
could provide new insights. The problem is however that such multifractal self-excited
process are extremely difficult to calibrate to empirical data. 
 
The scaling relation (\ref{rown12}) characterises in a novel way how the extreme losses interact with each other as we look at more and more extreme values, i.e. longer mean interevent times $\left< \tau\right>$. 

\section{Conclusions}

We have analyzed the probability density functions (PDFs) of interevent times between exceedance losses
in 15 empirical financial time series.
We showed that the PDFs can be accurately fitted by the scaling law (\ref{rown9}) with (\ref{rown10}),
which derives from the self-excited Hawkes conditional Poisson process with a
power law (Omori-Utsu) memory kernel.

Our work is based on previous works
analysing the PDFs of interevent times between successive earthquakes. We showed that the formalism used for earthquakes can be successfully used for modelling the PDFs of interevent times between successive market losses
exceeding a given threshold. This suggests that the observed approximate
universal scaling laws of interevent times may find a common origin in the mechanism
of self-excitation with long memory, which is the essential ingredient of the Hawkes process with
a power law memory kernel. Our calibration shows the strong non-Markovian nature 
of loss exceedances, and thus improve significantly on previous modelling efforts.

\section*{Acknowledgments} 
The work of M.J. was supported by a Swiss Government Excellence Scholarship.

\end{document}